\documentclass[reqno,a4paper,final,10pt]{amsart}
\usepackage[latin1]{inputenc}
\usepackage{amssymb,amstext,amsthm,eucal}
\usepackage{color}
\usepackage{array}
\usepackage[all]{xy}
\usepackage{hyperref}
\hypersetup{%
  pdftitle   = {On the non-existence of M-theory preons},
  pdfkeywords = {supergravity, supersymmetry, preons, quotients, waves}
  pdfauthor  = {José Figueroa-O'Farrill and Sunil Gadhia},
  pdfcreator = {\LaTeX\ with package \flqq hyperref\frqq}
}
\newcommand{\half}{\tfrac12}
\newcommand{\fg}{\mathfrak{g}}
\newcommand{\fgl}{\mathfrak{gl}}
\newcommand{\fh}{\mathfrak{h}}
\newcommand{\fs}{\mathfrak{s}}
\newcommand{\fso}{\mathfrak{so}}
\newcommand{\Cl}{\mathrm{C}\ell}
\newcommand{\Spin}{\mathrm{Spin}}
\newcommand{\Pin}{\mathrm{Pin}}
\ifx\Sp\UnDeFiNeD
\newcommand{\Sp}{\mathrm{Sp}}
\else
\renewcommand{\Sp}{\mathrm{Sp}}
\fi

\newcommand{\SO}{\mathrm{SO}}

\newcommand{\SL}{\mathrm{SL}}
\newcommand{\GL}{\mathrm{GL}}

\newcommand{\RR}{\mathbb{R}}
\newcommand{\NN}{\mathbb{N}}
\newcommand{\CC}{\mathbb{C}}
\newcommand{\ZZ}{\mathbb{Z}}

\renewcommand{\1}{\boldsymbol{1}}
\newcommand{\bzero}{\boldsymbol{0}}

\newcommand{\bM}{\boldsymbol{M}}
\newcommand{\be}{\boldsymbol{e}}
\newcommand{\bv}{\boldsymbol{v}}
\newcommand{\bz}{\boldsymbol{z}}
\DeclareMathOperator{\AdS}{AdS}

\DeclareMathOperator{\tr}{tr}
\DeclareMathOperator{\End}{End}
\DeclareMathOperator{\Mat}{Mat}
\DeclareMathOperator{\Ad}{Ad}

\DeclareMathOperator{\codim}{codim}
\DeclareMathOperator{\ind}{ind}
\DeclareMathOperator{\im}{Im}

\newtheorem{lemma}{Lemma}
\newcommand{\jparagraph}[1]{\paragraph{\bfseries #1}}
\newcommand{\MUNCH}[1]{\relax}

\allowdisplaybreaks[1]
\begin{document}
\title{M-theory preons cannot arise by quotients}
\author{José Figueroa-O'Farrill}
\author{Sunil Gadhia}
\address{Maxwell Institute and School of Mathematics, University of
  Edinburgh, UK}
\email{J.M.Figueroa@ed.ac.uk,S.Gadhia@sms.ed.ac.uk}
%\date{\today}
\begin{abstract}
  M-theory preons---solutions of eleven-dimensional supergravity
  preserving $31$ supersymmetries---have recently been shown to be
  locally maximally supersymmetric.  This implies that if preons exist
  they are quotients of maximally supersymmetric solutions.  In this
  paper we show that no such quotients exist.  This is achieved by
  reducing the problem to quotients by cyclic groups in the image of
  the exponential map, for which there already exists a partial
  classification, which is completed in the present paper.
\end{abstract}
\maketitle
\tableofcontents

\section{Introduction and outline}
\label{sec:intro}

M-theory preons, solutions of eleven-dimensional supergravity
preserving a fraction $\frac{31}{32}$ of the supersymmetry, were
conjectured in \cite{11dPreons} to be elementary constituents of other
BPS states.  They have been the subject of much recent work, reviewed
for instance in \cite{BandosReview}, until ultimately serious doubts
have been cast over their existence in \cite{NoMPreons}, where it is
shown that the gravitino connection of a preonic background is
necessarily flat, whence preons are locally maximally supersymmetric.
In \cite{PreonsConfinement} the earlier analogous result
\cite{NoIIBPreons} for type IIB supergravity was reinterpreted as
a ``confinement'' of supergravity preons; although a
dynamical mechanism which would be responsible for this confinement
has not been proposed.  In a similar vein one could say that M-theory
preons appear to be similarly confined.

The result in \cite{NoMPreons} implies that the universal cover of a
preonic background is maximally supersymmetric or, equivalently, that
the putative preonic background must be the quotient of a maximally
supersymmetric background by a discrete subgroup of the symmetry
group.  The aim of this paper is to show that if such a quotient
preserves a fraction $\nu \geq \frac{31}{32}$ of the supersymmetry
then it is in fact maximally supersymmetric, thus proving conclusively
that preons do not exist in eleven-dimensional supergravity or indeed
in any known supergravity theory with $32$ supercharges.

Let $(M,g,F)$ be a simply-connected maximally supersymmetric
background of $d{=}11$ supergravity and let $G$ denote the Lie group
of $F$-preserving isometries of the background.  We will let $\fg$
denote its Lie algebra.  Let $\Gamma < G$ be a discrete subgroup.
Then the quotient $M/\Gamma$ is a (possibly singular) background of
eleven-dimensional supergravity which is locally isometric to the
original background.  A natural question to ask is how much
supersymmetry will the quotient preserve.  Before we can ask this
question, we must ensure that $\Gamma$ can act on spinors.  Since
$\Gamma$ acts by isometries, it preserves the orthonormal frame
bundle.  The question is whether this action will lift to the spin
bundle.  We will assume it does.  Since $(M,g,F)$ is maximally
supersymmetric, the spinor bundle is trivialised by Killing spinors.
Let $K$ denote the space of Killing spinors, whence the spinor bundle
is isomorphic to $M \times K$, spinor fields are thus smoooth maps
$M\to K$ and the Killing spinors correspond to the constant maps.
Since $\Gamma$ also preserves $F$, it preserves the gravitino
connection and hence acts on the space of Killing spinors $K$.  The
converse is also true: if $\Gamma$ acts on $K$ then it will act on the
spinor bundle by combining the action on $K$ with that on $M$.  If
$\Gamma$ is contained in the identity component of $G$, then it does
act on $K$ and hence on spinors in general.  Indeed, the action of the
Lie algebra $\fg$ on $K$ is known explicitly, as it is a crucial
ingredient in the construction of the superalgebras associated to
these backgrouds.  Hence we may exponentiate this action to obtain an
action of the identity component of $G$.  Should the background
possess any ``discrete'' symmetries; that is, if $G$ has more than one
connected component, then we will assume that the group by which we
quotient does act on spinors.  The supersymmetry preserved by the
quotient background is again given by the Killing spinors in the
quotient.  Since the quotient is locally isometric to $M$, we may lift
this problem to $M$ and we see that Killing spinors of $M/\Gamma$
correspond to those Killing spinors on $M$ which are
$\Gamma$-invariant.  We will let $K^\Gamma$ denote the space of
$\Gamma$-invariant Killing spinors on $M$.  Similarly if $\gamma \in
\Gamma$, we will let $K^\gamma$ denote the space of Killing spinors of
$M$ which are left invariant by (the cyclic subgroup generated by)
$\gamma$.

In this paper we will show that if $\dim K^\Gamma \geq 31$, which we
call the \textbf{$31$-condition}, then $\dim K^\Gamma = 32$, whence it
is impossible to construct preonic backgrounds by quotients.  The
outline of the proof is the following.

We will let $R: G \to \GL(K)$ denote the action of $G$ on $K$.
Suppose that $\Gamma$ is such that $\dim K^\Gamma = 31$.  Then it is
plain that for some $\gamma \in \Gamma$, $\dim K^\gamma = 31$.
Indeed, were this not the case, then it would mean that either for all
$\gamma \in \Gamma$, $\dim K^\gamma = 32$, in which case also $\dim
K^\Gamma = 32$; or else for some $\gamma \in \Gamma$, $\dim K^\gamma
\leq 30$, whence also $\dim K^\Gamma \leq 30$.  It is therefore enough
to show that for no $\gamma \in G$, $\dim K^\gamma = 31$.  In the next
section we will prove this for $\gamma$ in the image of the
exponential map; that is, if $\gamma = \exp(X)$ for some $X \in \fg$,
then if $\dim K^\gamma \geq 31$ then $\dim K^\gamma = 32$, whence
$R(\gamma) = 1$.

Of course, in most of the groups $G$ under consideration, the
exponential map will not be surjective; however we will be able to
get around this problem as follows.  First let us consider the case
where $G$ is connected.

First of all we notice that every element $\gamma \in G$ is the
product of a finite number of elements in the image of the exponential
map.\footnote{In fact, it is shown in \cite{WuestnerSquare} that every
  element can be written as the product of at most \emph{two} elements
  in the image of the exponential map.}  This is because a
finite-dimensional Lie group is multiplicatively generated by any open
neighbourhood of the identity and we take one such neighbourhood to be
$\exp(S)$ for some open set $0 \in S \subset \fg$.  Denoting also by
$R:\fg \to \End(K)$ the action of $\fg$ on $K$, it will follow by
inspection of the relevant groups that $\tr R(X) = 0$ for all $X \in
\fg$, whence
\begin{equation*}
  \det R(\exp(X)) = \det e^{R(X)} = e^{\tr R(X)} = 1~,
\end{equation*}
whence $R : G \to \SL(K)$.
Now, if $\dim K^\gamma > 30$, $R(\gamma)$ must lie inside the subgroup
of $\SL(K)$ leaving invariant at least $31$ linearly independent
spinors.  In some basis, this is the subgroup
\begin{equation*}
\left\{
  \begin{pmatrix}
    I_{31} & \bv \\ \bzero^t & 1
  \end{pmatrix} \middle | \bv \in \RR^{31}\right\}~,
\end{equation*}
where $I_{31}$ is the $31\times 31$ identity matrix.  This means that
if for some power $k$, $\gamma^k$ lies in the image of the exponential
map, then
\begin{equation*}
  R(\gamma^k) = R(\gamma)^k = \begin{pmatrix} I_{31} & k \bv \\
    \bzero^t & 1 \end{pmatrix} = 1 \iff \bv = \bzero~,
\end{equation*}
so that $R(\gamma) = 1$.

The question is thus whether given $\gamma \in G$ some power of
$\gamma$ will lie in the image $E_G$ of the exponential map.  This
problem turns out to have some history, reviewed, for instance, in
\cite{Exponential}.  Given $\gamma \in G$, we define its \textbf{index
  (of exponentiality)} by
\begin{equation*}
  \ind (\gamma) =
  \begin{cases}
    \min \left\{ k \in \NN \middle| \gamma^k \in
      E_G\right\}~,& \text{should this exist}\\
    \infty& \text{otherwise.}
  \end{cases}
\end{equation*}
In all maximally supersymmetric backgrounds except for the wave, it is
known that every element of $G$ has finite index.  Thus the only case
which needs to be examined closely is the case where $G$ is the
solvable transvection group of a lorentzian symmetric space of
Cahen--Wallach \cite{CahenWallach} type.  The (simply-connected)
universal covering group $\widetilde G$ typically has elements with
infinite index.  However, $\widetilde G$ has no finite-dimensional
faithful representations and hence the finite-dimensional
representation $R: \widetilde G \to \SL(K)$ on Killing spinors will
factor through a $\ZZ$-quotient $\widehat G$ of $\widetilde G$ for
which it will be possible to prove that every element has finite
index.

Finally, let us consider the possibility that $G$ is not connected and
that $\Gamma$ is not contained in the identity component $G^0$ of $G$.
Let $\Gamma^0 = \Gamma \cap G^0$.  Then, assuming that $G$ has
finitely many components, $\Gamma/\Gamma^0$ is a finite group and the
representation $R: \Gamma \to \GL(K)$ factors through $\overline R :
\Gamma/\Gamma^0 \to \GL(K)$.  Since $K$ is a real representation,
$\det \overline R(\gamma) = \pm 1$.  If the determinant is $1$, then
we can again conclude that $\dim K^\Gamma = 32$.

To prove that $\overline R : \Gamma/\Gamma^0 \to \SL(K)$ we can argue
as follows.  Let $\gamma \in \Gamma\setminus\Gamma^0$. Because the
spacetime $M$ is connected, there is some element $\gamma_p \in G^0$
such that $h:=\gamma_p^{-1} \gamma$ fixes a point, say, $p\in M$.
(Indeed, for every $p\in M$ there will be some $\gamma_p\in G^0$ with
this property.  Namely, let $q = \gamma \cdot p$ and choose $\gamma_p
\in G^0$ such that $q = \gamma_p \cdot p$.  Such an element exists
because $M$ is connected and hence $G^0$ already acts transitively.)
Now the tangent map $h_* : T_pM \to T_pM$ defines an orthogonal
transformation on $T_pM$ which, by hypothesis, lifts to an action on
the Killing spinors, and which is induced by restriction from the
action of the $\Pin$ group.  If the spin lift of $h_*$ acts with unit
determinant, then from the fact that $\gamma_p$ does so as well, it
follows that so will $\gamma$.  It is then a matter of verifying that
the $\Pin$ group acts with unit determinant on the relevant spinor
representation.

The paper is organised as follows.  In Section \ref{sec:Mmax} we check
that no quotient by a cyclic subgroup of symmetries in the image of
the exponential map preserves $31$ supersymmetries.  The result for
flat space follows (at least implicitly) from results in
\cite{FigSimFlat}.  It is discussed in §\ref{sec:Mflat} for
completeness and because it is the simplest setting in which to
present what we call the \emph{even-multiplicity} argument, which is
used throughout the paper.  The rest of the section is taken by the
Freund--Rubin backgrounds, discussed in §\ref{sec:Mfreund-rubin} using
the notation of \cite{FigSimAdS}, as well as the maximally
supersymmetric wave, discussed in §\ref{sec:Mwave} for the first time.
In Section \ref{sec:exp} we investigate the surjectivity properties of
the exponential map for the symmetry groups of the relevant vacua.  A
large part of the discussion is devoted to showing that every element
of the symmetry group of the maximally supersymmetric wave has finite
index.  Finally in Section \ref{sec:other} we briefly discuss other
supergravity theories and conclude that there are no preonic
supergravity backgrounds in any known supergravity theory with $32$
supercharges.

\section{Cyclic quotients of M-theory vacua}
\label{sec:Mmax}

In this section we review the possible quotients of the
simply-connected maximally supersymmetric eleven-dimensional
supergravity backgrounds by the action of the subgroup generated by an
element $\gamma = \exp(X)$ for $X\in\fg$, the Lie algebra of
$F$-preserving isometries of the background.  We will show that no
quotient preserves exactly a fraction $\frac{31}{32}$ of the
supersymmetry.

The method of classification has been explained before in a series of
papers \cite{FigSimFlat,FigSimBranes,FigSimGrav,FigSimAdS} to where we
refer the reader interested in the details.  The basic idea 
is to study the orbit decomposition of the symmetry Lie algebra $\fg$
under the adjoint action of the Lie group $G$.  Fixing a
representative from each orbit, we may then study its action on the
Killing spinors.  The determination of the adjoint orbits has already
been done for all backgrounds but the maximally supersymmetric wave
\cite{KG,FOPflux}, which is the subject of the last subsection.
The action on the Killing spinors is a group-theoretical problem which
we address in this section.

\subsection{Minkowski background}
\label{sec:Mflat}

The quotients of $\RR^{1,10}$ by continuous cyclic subgroups have been
discussed in \cite{FigSimFlat} and the results on discrete quotients
follow easily from these.  First of all we notice that translations
act trivially on spinors, hence the amount of supersymmetry which is
preserved by a group element $\gamma \in \Spin(10,1)\ltimes \RR^{11}$
is governed by its projection onto $\Spin(10,1)$.  Hence from now on
we will consider $\gamma \in \Spin(10,1)$ of the form $\gamma =
\exp(X)$, for some $X \in \fso(10,1)$.  As explained in
\cite{FigSimFlat}, there are three possible maximal conjugacy classes
of such $X$, depending on the causal type of the vector they leave
fixed infinitesimally:
\begin{enumerate}
\item $X = \theta_1 \be_{12} + \theta_2 \be_{34} + \theta_3 \be_{56} +
  \theta_4 \be_{78} + \theta_5 \be_{9\natural}$~,
\item $X = \theta_1 \be_{12} + \theta_2 \be_{34} + \theta_3 \be_{56} +
  \theta_4 \be_{78} + \beta \be_{09}$~, and
\item $X = \theta_1 \be_{12} + \theta_2 \be_{34} + \theta_3 \be_{56} +
  \theta_4 \be_{78} + \be_{+9}$~,
\end{enumerate}
where $\be_{ij} := \be_i \wedge \be_j \in \Lambda^2\RR^{10,1} \cong
\fso(10,1)$, with $\be_i$ a pseudo-orthonormal basis for $\RR^{10,1}$,
and $\be_+ := \be_0 + \be_{\natural}$, where as usual $\natural$
stands for $10$.  The Killing spinors of the Minkowski background are
isomorphic, as a representation of $\Spin(10,1)$, with the spinor
module $\Delta^{10,1}$, which is real and $32$-dimensional.  We find
it convenient to work in the Clifford algebra $\Cl(10,1)$ which
contains the relevant spin group.  As an associative algebra,
$\Cl(10,1) \cong \Mat_{32}(\CC)$, whence it has a unique irreducible
module $W$, which is complex and $32$-dimensional.  As a
representation of $\Spin(10,1)$ it is the complexification of the
spinor representation.

The first two cases in the above list are a special case of the
following set-up.

\subsubsection{The even-multiplicity argument}
\label{sec:even-multiplicity}

Let $I_a$, for $a=1,\dots,N$, be commuting real ($I_a^2=1$) or complex
($I_a^2=-1$) structures and consider $R(\gamma):=\exp(\sum_a \half
\theta_a I_a)$ acting on a complex vector space $W$ of dimension $2^N$.
Since the $I_a$ are commuting, we may diagonalise them 
simultaneously and decompose
\begin{equation*}
  W = \bigoplus_{(\sigma_1,\dots,\sigma_N)\in\ZZ_2^N}
  W_{\sigma_1\dots\sigma_N}
\end{equation*}
where on each one-dimensional $W_{\sigma_1\dots\sigma_N}$, $\gamma$
acts by $e^{\sum_a \varepsilon_a \sigma_a \theta_a/2}$, where
\begin{equation*}
  \varepsilon_a =
  \begin{cases}
    1~, &\text{if $I_a$ is a real structure,}\\
    i~, &\text{if $I_a$ is a complex structure.}
  \end{cases}
\end{equation*}

We now observe that if $R(\gamma)$ acts as the identity on some
$W_{\sigma_1\dots\sigma_N}$, it also acts as the identity on
$W_{\bar\sigma_1\dots\bar\sigma_N}$, where $\bar\sigma_a = -
\sigma_a$.  This means that the subspace $W^\gamma$ of
$\gamma$-invariants has even complex dimension.  Being a real
representation, it is the complexification of an even-dimensional real
subspace.  Therefore it cannot be odd-dimensional.

The preceding argument is quite general and will be applied
below also in the Freund--Rubin and wave backgrounds.  We will refer
to it as the \textbf{even-multiplicity} argument.

\jparagraph{Cases 1 and 2}

In these cases, and in the notation of the preceding discussion, $N=5$
and $W$ is the complexification of $\Delta^{10,1}$, whereas the $I_a$
are the images in the Clifford algebra of the infinitesimal rotations
$\be_{12},\be_{34}, \be_{56}, \be_{78}, \be_{9\natural}$ or the
infinitesimal boost $\be_{09}$ in $\fso(10,1)$.  Applying the
even-multiplicity argument, we see that the $\gamma$-invariant
subspace is even-dimensional and hence if its dimension is $>30$, it
must be $32$.

\jparagraph{Case 3}

In this case, the group element is $R(\gamma) = \exp(N + \sum_a\half
\theta_a I_a)$, where $N$ is the image of the infinitesimal null
rotation $\be_{09}-\be_{9\natural} \in \fso(10,1)$ under the spin
representation.  It follows that $N^2=0$ in the Clifford algebra,
whence $\exp(N) = 1 + N$ in the spin group.  We are after the
dimension of the subspace of $W$ consisting of $\psi \in W$ satisfying
\begin{equation}
  \label{eq:N-inv}
  R(\gamma) \psi = \exp(\sum_a\half \theta_a I_a) (1 + N) \psi =
  \psi~.
\end{equation}
Let us break up $\psi = \psi_+ + \psi_-$ according to 
\begin{equation*}
  V = V_+ \oplus V_-~,
\end{equation*}
where $V_\pm = \ker (\be_0 \pm \be_\natural)$, understood as Clifford
product.  Clearly, $\ker N = \im N  = V_+$.  Equation~(\ref{eq:N-inv})
becomes
\begin{equation*}
  \exp(\sum_a\half \theta_a I_a) (\psi_+ + \psi_- + N\psi_-) = \psi_+ + \psi_-~,
\end{equation*}
which in turn breaks up into two equations
\begin{equation*}
  \exp(\sum_a\half \theta_a I_a) \psi_- = \psi_- \qquad\text{and}\qquad
  \exp(\sum_a\half \theta_a I_a) (\psi_+ + N\psi_-) = \psi_+~.
\end{equation*}
The even-multiplicity argument says that the invariant space is even
dimensional, hence for the $31$-condition to hold, the first equation
forces $\exp(\sum_a\half \theta_a I_a) = 1$.  The second equation then becomes
$N\psi_- = 0$, which means that $\psi_-=0$.  Therefore the most
supersymmetry that such a quotient preserves is precisely one half.

\subsection{Freund--Rubin backgrounds}
\label{sec:Mfreund-rubin}

The one-parameter quotients of the Freund--Rubin backgrounds, $\AdS_4
\times S^7$ and $\AdS_7 \times S^4$, have been discussed in
\cite{FigSimAdS,MaddenRoss} and the discrete quotients by subgroups in
the image of the exponential map have been discussed in \cite{FOMRS}.
The emphasis in those papers were on quotients which preserve causal
regularity. Among such quotients, those by the subgroup generated by
(a power of) the generator of the centre of the isometry group of
$\AdS$ preserve all the supersymmetry.  The resulting space is a
finite cover of the hyperboloid model for $\AdS$ (times the sphere)
and admits closed time-like curves. The only other quotients among
them preserving more than half of the supersymmetry are described in
detail in \cite[Appendix B]{FMPHom} and we will not repeat the
calculation here.  They preserve fractions $\frac34$ and $\frac9{16}$
of the supersymmetry.  In summary, there are no such quotients
preserving exactly $\frac{31}{32}$ of the supersymmetry.

For the present purposes, however, the restriction to causally regular
quotients is not desirable and we must therefore revisit the
classifications in \cite{FigSimAdS} and study how much supersymmetry
is preserved in each case.  Every element in the image of the
exponential group takes the form $\exp(X)$ for some $X \in \fg$ in the
Lie algebra of the symmetry group of the background.  Two elements
$X,Y \in \fg$ which lie in the same adjoint orbit are equivalent for
our purposes since their action on Killing spinors will be related by
conjugation and hence, in particular, will leave the same number of
Killing spinors invariant.  The adjoint orbits have been classified in
\cite{FigSimAdS}, to whose notation we will adhere in what follows.

\subsubsection{$\AdS_4 \times S^7$}
\label{sec:ads4s7}

In this case, the Lie algebra of symmetries is $\fso(3,2) \oplus
\fso(8)$ and its action on the Killing spinors is given by the tensor
product representation $\Delta^{3,2} \otimes \Delta_-^8$, where
$\Delta^{3,2}$ is the real $4$-dimensional spin representation of
$\fso(3,2)$ and $\Delta_-^8$ is the real $8$-dimensional half-spin
representation of $\fso(8)$ consisting of negative chirality spinors.
The typical element $X \in \fg$ decomposes as $X_A + X_S$, with $X_A
\in \fso(3,2)$ and $X_S \in \fso(8)$.  Every element $X_S \in \fso(8)$
belongs to some Cartan subalgebra and these are all conjugate.
Therefore we may always bring $X_S$ to the form $\theta_1 R_{12} +
\theta_2 R_{34} + \theta_3 R_{56} + \theta_4 R_{78}$, where $R_{ij}$
is the element of $\fso(8) = \fso(\RR^8)$ which generates rotations in
the $ij$-plane.  In contrast there are $15$ possible choices for
$X_A$, which are listed in \cite[§4.2.1]{FigSimAdS}.  It is natural
for the present purposes to treat some of these cases together, which
explains the subdivision below.

We will perform our calculations in the Clifford algebra $\Cl(11,2)$,
which contains $G=\Spin(3,2) \times \Spin(8)$, the spin group in
question.  As an associative algebra, $\Cl(11,2) \cong \Mat_{64}(\CC)$
and hence has a unique irreducible module $W$, which is
$64$-dimensional and complex and which decomposes under $G$ into the
direct sum of $32$-dimensional complex subrepresentations (with a real
structure) corresponding to $\Delta^{3,2} \otimes \Delta_+^8$ and
$\Delta^{3,2} \otimes \Delta_-^8$.  We are interested in the real
representation whose complexification is $V = \Delta^{3,2} \otimes
\Delta_-^8$.

We notice that in many of the cases below we will be able to apply the
even-multiplicity argument.  There is only one subtlety and that is
that we are interested not in $W$ but on a subspace $V$ determined by
some chirality condition.  In the notation of
§~\ref{sec:even-multiplicity}, we have $N=6$ and
\begin{equation*}
  V = \bigoplus_{\substack{(\sigma_1,\dots,\sigma_6)\in\ZZ_2^6\\
    \sigma_3\sigma_4\sigma_5\sigma_6=-1}} W_{\sigma_1\dots\sigma_6}~,
\end{equation*}
where the constraints on the signs comes from the chirality condition
for $\fso(8)$.  To apply the even-multiplicity argument we need to
check that if $W_{\sigma_1\dots\sigma_6} \subset V$, then also
$W_{\bar\sigma_1\dots\bar\sigma_6} \subset V$, for $\bar\sigma_a = -
\sigma_a$.  This is clear, though, by definition of the constraints
defining $V$.

In summary, the even-multiplicity argument applies to all cases where
$X_A$ consists only of $2\times 2$ blocks in the language of
\cite{FigSimAdS}; that is, to case 1, 2, 4, 10, 11, and 12.  The
remaining cases contain blocks of higher dimension and must be
analysed separately.   Cases 3,5,14 and 15 are virtually identical to
Case 3 in §\ref{sec:Mflat} and will not be discussed further.

\jparagraph{Cases 6, 7 and 8}

These cases are very similar and are defined by the $\fso(3,2)$
component, which can take one of the following forms
\begin{itemize}
\item $X^{(6)}_A = -\be_{12} - \be_{13} + \be_{24} + \be_{34}$,
\item $X^{(7)}_A = -\be_{12} - \be_{13} + \be_{24} + \be_{34} + \beta
  (\be_{14} - \be_{23})$, and
\item $X^{(8)}_A = -\be_{12} - \be_{13} + \be_{24} + \be_{34} + \theta
  (\be_{12} + \be_{34})$.
\end{itemize}
We will focus on $X_A^{(8)}$, which will specialise trivially to
$X_A^{(6)}$ and leave $X_A^{(7)}$ as a very similar exercise.
Let $N + \theta T$ denote the image of $X^{(8)}_A$ in the Clifford
algebra, with $N = (\be_2 + \be_3)(\be_1+\be_4)$.  It follows that $N
T = T N = 0$ and that $N^2 = 0$.  Therefore the group element is given
by
\begin{equation*}
  \exp(N + \theta T + \sum_{a>2} \theta_a I_a) = \exp(\sum_a
  \theta_a I_a) (1 + N)~,
\end{equation*}
where we have put $\theta_1 = \theta_2 = \theta$.  We find it
convenient to decompose $V$ into four eight-dimensional subspaces
\begin{equation*}
  V = V_{++} \oplus V_{+-} \oplus V_{-+} \oplus V_{--}~,
\end{equation*}
where $V_{\pm\pm} = \ker (\be_2 \pm \be_3) \cap \ker (\be_1 \pm
\be_4)$ with uncorrelated signs.  Then $N$ acts trivially except on
$V_{--}$ where it defines a map $V_{--} \to V_{++}$.
Let us write the invariance condition as
\begin{equation*}
  s \exp(N + T) \psi = \psi~,
\end{equation*}
where $s \in \Spin(8)$.  Expanding the exponentials, using that $T$
and $N$ commute and that $N^2 = 0$, we arrive at
\begin{equation*}
  s \exp(T) (1 + N) \psi = e \exp(T) \psi + s\exp(T) N \psi = \psi~.
\end{equation*}
The term in the image of $N$ is in $V_{++}$ and because $NT = 0$, 
$\im T \cap V_{--} = \varnothing$, whence focusing on the
$V_{--}$ part of this equation, we find that
\begin{equation*}
  s \psi_{--} = \psi_{--}~.
\end{equation*}
The $31$-condition forces $s=1$, and that means that any element of
$\Spin(3,2)$, in particular $\exp(T+N)$, acts with multiplicity $8$ on
$V$.  Hence the dimension of the invariant subspace is a multiple of
$8$, which cannot therefore be equal to $31$.

\jparagraph{Case 9}

In this case $X_A = \varphi (\be_{12} - \be_{34}) + \beta
(\be_{14}-\be_{23}) \in \fso(3,2)$.  The corresponding element
$\exp(X_A)$ in $\Spin(3,2) \subset \Cl(3,2)$ is given by
$\exp(\half\varphi A + \half\beta B)$, where $A$ and $B$ are the
images of $\be_{12}-\be_{34}$ and $\be_{14}-\be_{23}$, respectively,
in the Clifford algebra.  It is easy to check that $AB = BA = 0$,
whereas $A^2 = -P_+$, $A^3= -A$, and similarly $B^2 = P_-$ and $B^3 =
B$, where $P_\pm = \half (1 \pm \be_{1234})$.  Decompose $V = V_+
\oplus V_-$, where $V_\pm = \im P_\pm$.  Letting $s = \exp(X_S) \in
\Spin(8)$, we want to determine the dimension of the subspace of
spinors $\psi$ satisfying
\begin{equation}
  \label{eq:9-inv}
  s \exp(\half \varphi A) \exp(\half \beta B) \psi = \psi~.
\end{equation}
Decomposing $\psi = \psi_+ + \psi_-$, with $\psi_\pm = P_\pm \psi$, we
find that the invariance equation (\ref{eq:9-inv}) breaks up into two
equations
\begin{equation*}
  s\exp(\half\varphi A)\psi_+ = \psi_+ \qquad\text{and}\qquad
  s\exp(\half\beta B)\psi_- = \psi_-~.
\end{equation*}
This latter equation implies that $\beta =0$ for more than half of the
supersymmetry to be preserved.  The $31$-condition then forces $s=1$,
since $s$ acts with even multiplicities.  This in turn implies that
any element in $\Spin(3,2)$ acts with multiplicity $8$, whence the
dimension of the invariant subspace is a multiple of $8$ and therefore
cannot be equal to $31$.

\jparagraph{Case 13}

Finally, we consider the case where $X_A= \be_{12} + \be_{13} +
\be_{15} - \be_{24} - \be_{34} - \be_{45}$.  A calculation in the
Clifford algebra shows that
\begin{equation}
  \label{eq:13-X_A}
  \exp(X_A) = 1 + (\be_1 + \be_4) (\be_2 + \be_3) - \be_5 (\be_1 -
  \be_4) - 2 \be_{145}(\be_2 + \be_3) - \tfrac23 (\be_1 - \be_4) (\be_2
  + \be_3)~.
\end{equation}
Letting $V_{\pm\pm} = \ker (\be_1 \pm \be_4) \cap \ker (\be_2 \pm
\be_3)$ with uncorrelated signs, $V$ decomposes as the direct sum
\begin{equation*}
  V = V_{++} \oplus V_{+-} \oplus V_{-+} \oplus V_{--}~.
\end{equation*}
We want to find the dimension of
the subspace of $\psi$ satisfying the equation
\begin{equation}
  \label{eq:13-inv}
  s \exp(X_A) \psi = \psi~,
\end{equation}
where we have let $s = \exp(X_S) \in \Spin(8)$.  Inspection of
equation (\ref{eq:13-X_A}) reveals that the $V_{+-}$ component of
equation (\ref{eq:13-inv}) is simply
\begin{equation*}
  s \psi_{+-} = \psi_{+-}~,
\end{equation*}
which, since $s$ acts with even multiplicities, forces $s=1$ after
invoking the $31$-condition.  This then implies that any element of
$\Spin(3,2)$ acts with multiplicity $8$ and in particular that the
dimension of the subspace of invariants is a multiple of $8$ and
cannot therefore be equal to $31$.

\subsubsection{$\AdS_7 \times S^4$}
\label{sec:ads7s4}

The Lie algebra of symmetries is now $\fso(6,2) \oplus \fso(5)$ and
its action on the Killing spinors is given by the underlying real
representation of the tensor product representation $\Delta^{6,2}_-
\otimes \Delta^5$, where $\Delta^{6,2}_-$ is the quaternionic
representation of $\fso(6,2)$ consisting of negative chirality spinors
and having complex dimension $8$, and $\Delta^5$ is the quaternionic
spin representation of $\fso(5)$, which has complex dimension $4$.
The typical element $X \in \fg$ again decomposes as $X_A + X_S$, with
$X_A \in \fso(6,2)$ and $X_S \in \fso(5)$.  As before, every element
$X_S \in \fso(5)$ belongs to some Cartan subalgebra and may be brought
to the form $\theta_1 R_{12} + \theta_2 R_{34}$.  In contrast now
there are $39$ possible choices for $X_A$, which are listed in
\cite[§4.4.1]{FigSimAdS}.  It is again natural for the present
purposes to treat some of these cases together, which explains the
following subdivision.

The general argument at the start of the previous subsection can again
be deployed to discard the cases with only $2\times 2$ blocks; that
is, rotations or boosts.  These are the cases 1, 2, 4, 10, 11, 12, 16,
24, 25, 26, 30, 38 and 39 in \cite[§4.4.1]{FigSimAdS}.

Many of the remaining cases already appeared in our discussion of
$\AdS_4 \times S^7$ and we will not repeat the arguments here for they
are virtually identical to the ones above.  These are cases 3, 5, 14,
15, 17, 28, 29 and 31 (which are similar to cases 3, 5, 14 and 15
above); cases 6, 7, 8, 19, 20, 21, 33, 34 and 35 (similar to cases
6,7,8 above); cases 9, 22 and 36 (similar to case 9 above); and cases
13 and 27 (similar to case 13 above).  The remaining cases can be
subdivided as follows.

\jparagraph{Cases 23 and 37}

This case involves a double null rotation.  The corresponding $X_A =
N_1 + N_2 + \theta \be_{78}$, where $N_1=(\be_1+\be_4)\be_3$ and $N_2
= (\be_2 + \be_6)\be_5$, already written in the Clifford algebra.
Notice that $N_1^2=N_2^2= 0$ and that $N_1 N_2 = N_2 N_1$, whence
$\exp(N_1+N_2) = (1+N_1)(1+N_2)$.  We decompose the space $V$ of
complexified Killing spinors as
\begin{equation*}
  V = V_{++} \oplus V_{+-} \oplus V_{-+} \oplus V_{--}~,
\end{equation*}
where $V_{\pm\pm} = \ker (\be_1 \pm \be_4) \cap \ker (\be_2 \pm
\be_6)$ with uncorrelated signs.  A spinor $\psi \in V$ is invariant
if it obeys
\begin{equation*}
  s (1+N_1)(1+N_2) \psi = \psi~,
\end{equation*}
where $s = R\left(\exp(\theta \be_{78})\exp(X_S)\right)$.  The
$V_{--}$ component of this equation is
\begin{equation*}
  s \psi_{--} = \psi_{--}~,
\end{equation*}
whence, since $s$ acts with even multiplicities, the $31$-condition
forces $s=1$.  This means that the $\Spin(5)$ part of $\gamma$ acts
trivially and so $\gamma$ acts with multiplicity $4$, whence the
dimension  of the space of invariants is a multiple of $4$, which if
$>30$ must therefore be equal to $32$.

\jparagraph{Cases 18 and 32}

In this case,
\begin{equation*}
  X_A = \be_{15} - \be_{35} + \be_{26} - \be_{46} + \varphi (
  -\be_{12} + \be_{34} + \be_{56} ) + \theta \be_{78}~.  
\end{equation*}
We concentrate on the exceptional $6\times 6$ block
\begin{equation*}
  A + \varphi B = \be_{15} - \be_{35} + \be_{26} - \be_{46} + \varphi
  (-\be_{12} + \be_{34} + \be_{56} )~.
\end{equation*}
In the Clifford algebra, we find that $AB=BA$, whence in the spin
group, $\exp(A+\varphi B) = \exp(A) \exp(\varphi B)$.  Therefore we
have
\begin{equation*}
  \exp(X) = \exp(A) \exp(\varphi B + \theta \be_{78} + X_S)~.
\end{equation*}
The second exponential is a semisimple element of the form
$\exp(\sum_a \theta_a I_a)$ for commuting complex structures $I_a$ and
therefore acts with even multiplicities.  Let $W = \Delta^{6,2}\otimes
\Delta^5$ be the complex $64$-dimensional irreducible module of the
Clifford algebra $\Cl(11,2)$.  Under the action of the 
six complex structures $I_a$ it decomposes into a direct sum of
one-dimensional subspaces
\begin{equation*}
  W = \bigoplus_{(\sigma_1,\dots,\sigma_6)\in\ZZ_2^N}
  W_{\sigma_1\dots\sigma_6}~,
\end{equation*}
whereas the subspace $V=\Delta^{6,2}_-\otimes\Delta^5$ of complexified
Killing spinors decomposes as
\begin{equation*}
  V = \bigoplus_{\substack{(\sigma_1,\dots,\sigma_6)\in\ZZ_2^6\\
    \sigma_1\sigma_2\sigma_3\sigma_4=-1}} W_{\sigma_1\dots\sigma_6}~,
\end{equation*}
where the constraints on the signs comes from the chirality condition
for $\fso(6,2)$.  The semisimple element $s = \exp(\varphi B + \theta
\be_{78} + X_S)$ preserves each subspace $W_{\sigma_1\dots\sigma_6}
\subset V$ acting on it by the scalar
\begin{equation*}
 e^{i\left(\varphi(-\sigma_1+\sigma_2+\sigma_3) + \theta\sigma_4 +
    \theta_5 \sigma_5 + \theta_6 \sigma_6\right)/2}~. 
\end{equation*}
The spectrum of $s$ on $V$ is therefore given by
\begin{itemize}
\item $e^{\pm i(\varphi-\theta + \sigma_5\theta_5 +
    \sigma_6\theta_6)/2}$ each with multiplicity $2$---a total of
  $16$;
\item $e^{\pm i(\varphi+\theta + \sigma_5\theta_5 + \sigma_6\theta_6)/2}$
each with multiplicity $1$---a total of $8$; and
\item $e^{\pm i(3\varphi+\theta + \sigma_5\theta_5 +
    \sigma_6\theta_6)/2}$ each with multiplicity $1$---a total of
  $8$.
\end{itemize}
Since $A$ commutes with $s$, $\exp(A)$ preserves each of these
eigenspaces.  Since $A^3 = 0$, the only possible eigenvalue of $e^A$
is $1$.  This means that the component of an invariant spinor $\psi$
belonging to any one of the above eigenspaces of $s$ must have
eigenvalue $1$.  The dimension of the eigenspace of $s$ with
eigenvalue $1$ is even, hence the $31$-condition says that this
eigenspace must be $32$-dimensional, or in other words that $s=1$,
whence the $\Spin(5)$ part of $\gamma$ acts trivially.  This implies
that $\gamma$ acts with multiplicity $4$, whence the dimension of the
space of invariant Killing spinors must be a multiple of 4 and hence,
if $>30$ it must be $32$.

\subsection{Maximally supersymmetric wave}
\label{sec:Mwave}

The cyclic quotients of the maximally supersymmetric wave have not
been worked out before, and we do so here.  We will base our
discussion of the maximally supersymmetric wave \cite{KG} on the paper
\cite{FOPflux}.  In particular, the geometry is that of a lorentzian
symmetric space $G/H$, where the transvection group $G$ and the
isotropy subgroup $H$ are described as follows.  Let $\fg$ be the
$20$-dimensional Lie algebra with basis
$(\be_\pm,\be_i,\be_i^*)$, for $i=1,\ldots,9$, and nonzero
brackets
\begin{equation*}
  [\be_-,\be_i]=\be_i^* \qquad [\be_-,\be_i^*] = -\lambda_i^2 \be_i\qquad
  [\be_i^*,\be_j]=-\lambda_i^2 \delta_{ij}\be_+~,
\end{equation*}
where
\begin{equation}
  \label{eq:CW11}
  \lambda_i =
  \begin{cases}
    \frac{\mu}3~,& i=1,2,3\\
    \frac{\mu}6~,& i=4,\ldots,6
  \end{cases}\qquad\text{and $\mu\neq 0$.}
\end{equation}
Let $\fh$ denote the abelian Lie subalgebra spanned
by the $\{\be_i^*\}$ and let $H<G$ denote the corresponding Lie subgroup.
The obvious subgroup $\SO(3) \times \SO(6) < \SO(9)$ acts as
automorphisms on $\fg$ preserving $\fh$ and hence acts as isometries
on $G/H$.  Moreover $S := G \rtimes \left(\SO(3)\times\SO(6)\right)$
preserves the four-form flux, hence it is also the symmetry group of
the background.

Let $\fs = \fg \rtimes \left(\fso(3) \oplus \fso(6)\right)$ denote the
Lie algebra of $S$.  Let us examine the possibility of bringing
$X\in\fs$ to a normal form via the adjoint action of $S$.  To this
end, the following expressions are useful:
\begin{equation}
  \label{eq:expad}
  \begin{aligned}[m]
    \Ad\left(e^{t \be_-}\right)\be_i &= \cos(\lambda_i t) \be_i + \frac{\sin(\lambda_i t)}{\lambda_i} \be_i^*\\
    \Ad\left(e^{t \be_-}\right)\be_i^* &= \cos(\lambda_i t) \be_i^* + \lambda_i \sin(\lambda_i t) \be_i\\
    \Ad\left(e^{t \be_i}\right)\be_- &= \be_- - t \be_i^* - \half t^2 \lambda_i^2 \be_+\\
  \end{aligned}\quad
  \begin{aligned}[m]
    \Ad\left(e^{t \be_i} \right)\be_j^* &= \be_j^* - t\lambda_i^2 \delta_{ij} \be_+\\
    \Ad\left(e^{t e^*_i} \right)\be_- &= \be_- + \lambda_i^2 t \be_i - \half t^2 \lambda_i^4 \be_+\\
    \Ad\left(e^{t e^*_i} \right)\be_j &= \be_j - t\lambda_i^2 \delta_{ij}\be_+~,
  \end{aligned}
\end{equation}
whereas the adjoint action of $\SO(3) \times \SO(6)$ is the obvious
one.  Without loss of generality we may assume that the $\left(\fso(3) \oplus
\fso(6)\right)$-component of $X$ lies in the Cartan subalgebra spanned
by $\{\bM_{12},\bM_{45},\bM_{67},\bM_{89}\}$, while still retaining
the freedom of acting with the associated maximal torus $T$, say.

We must now distinguish two cases, depending on whether the component
of $X$ along $\be_-$ does or does not vanish.  If it does not vanish,
then from equation \eqref{eq:expad} it follows that we may set the
$\be_i$-components of $X$ equal to zero by acting with
$\Ad\left(e^{t\be^*_i}\right)$.   Acting with
$\Ad\left(e^{t\be_3}\right)$ we may shift the $\be^*_3$-component to
zero.  Acting further with $T$ we may rotate in the $(\be^*_1,\be^*_2)$,
$(\be^*_4,\be^*_5)$, $(\be^*_6,\be^*_7)$ and $(\be^*_8,\be^*_9)$
planes to set the $\be^*_i$-components to zero for $i=2,5,7,9$.  This
brings $X$ to the
following form
\begin{equation}
  \label{eq:X1}
  X = v^+\be_+ + v^-\be_- + v^1\be^*_1 + v^4 \be^*_4 + v^6\be^*_6 +
  v^8\be^*_8 + \theta^1 \bM_{12} + \theta^2 \bM_{45} + \theta^3
  \bM_{67} + \theta^4 \bM_{89}~,
\end{equation}
where $v^-\neq 0$.

If the $\be_-$-component of $X$ vanishes, then we may set the
components along some of the $\be_i$ to zero, but not much else.  This
brings $X$ to the form
\begin{multline}
  \label{eq:X2}
  X = v^+\be_+ + \sum_i v^i\be^*_i + w^1 \be_1 + w^4 \be_4 + w^6 \be_6
  + w^8\be_8\\
  + \theta^1 \bM_{12} + \theta^2 \bM_{45} + \theta^3 \bM_{67} +
  \theta^4 \bM_{89}~.
\end{multline}

The action of $X$ on the space $K$ of Killing spinors can be read off
from the calculation of the superalgebra in \cite[§6]{FOPflux}.  Let
$R: \fs \to \End(K)$ denote the representation, we find
\begin{equation*}
  \begin{aligned}[m]
    R(\be_i) &= -\half \lambda_i I \Gamma_i \Gamma_+\\
    R(\be_i^*) &= - \half \lambda_i^2 \Gamma_i \Gamma_+
  \end{aligned}\qquad\qquad
  \begin{aligned}[m]
    R(\be_-) &= -\frac{\mu}{4} I \Pi_+ - \frac{\mu}{12} I \Pi_-\\
    R(\bM_{ij}) &= \half \Gamma_{ij}~,
  \end{aligned}
\end{equation*}
where $\{\Gamma_+,\Gamma_-,\Gamma_i\}$ are the $\Cl(1,9)$ gamma
matrices in a Witt basis, $I=\Gamma_{123}$ and $\Pi_\pm = \half
\Gamma_\pm \Gamma_\mp$ are the projectors onto $\ker \Gamma_\pm$ along
$\ker\Gamma_\mp$.  We follow the conventions in \cite{FOPflux}, so
that $\Gamma_+ \Gamma_- + \Gamma_- \Gamma_+ = 2 \1$ and $\Gamma_i^2 =
\1$.  In particular, whilst $\{R(\be_-),R(\bM_{ij})\}$ are semisimple,
$\{R(\be_i),R(\be_i^*)\}$ are nilpotent.  This means that for $X \in
\fs$, we may decompose $R(X) = R(X)_S + R(X)_N$ into semisimple and
nilpotent parts.  Exponentiating and using the BCH formula, we find
\begin{equation*}
  R(\gamma) := e^{R(X)} = e^{R(X)_S + R(X)_N} = g_S g_N =
  g'_N g_S~,
\end{equation*}
where $g_S=e^{R(X)_S}$ and $g_N$ and $g'_N$ are exponentials of
nilpotent endomorphisms.  In particular, given the nature of the
nilpotent endomorphisms in the image of $R$, we know that $g_N = \1 +
\alpha \Gamma_+$ and similarly for $g'_N$.

Before specialising to a particular form of $X$, let us make some
general remarks about the amount of supersymmetry preserved by
$R(\gamma)$.  The space $K$ of Killing spinors decomposes as $K = K_+
\oplus K_-$, where $K_\pm = K \cap \ker \Gamma_\pm$.  Let $\psi =
\psi_+ + \psi_-$, with $\psi_\pm \in K_\pm$, be a $\gamma$-invariant
Killing spinor, so that $R(\gamma)\psi = g_S g_N \psi = \psi$.
Decomposing this equation, and using $g_N = \1 + \alpha \Gamma_+$, we
find
\begin{equation*}
  g_S g_N (\psi_+ + \psi_-) = g_S (\psi_+ + \psi_- + \alpha \Gamma_+
  \psi_-) = \psi_+ + \psi_-~.
\end{equation*}
Since $g_S$ respects the decomposition $K_+ \oplus K_-$, we see that,
in particular, $g_S \psi_- = \psi_-$.  We would like to estimate how
big a subspace of $K_-$ this is.

Let $K^0 \subset K$ denote the subspace of $R(\gamma)$-invariant Killing
spinors and let $K^0_\pm = K^0 \cap K_\pm$.  Then letting $K^0 + K_-$
denote the subspace of $K$ generated by $K^0$ and $K_-$, we have the
fundamental identity
\begin{equation*}
  \dim(K^0 + K_-) - \dim K^0  = \dim K_- - \dim(K^0 \cap K_-)~.
\end{equation*}
Since $\dim(K^0 + K_-) - \dim K^0 \leq \codim(K^0\subset K)$, we
arrive at
\begin{equation*}
  \codim(K_-^0\subset K_-) \leq \codim (K^0\subset K)~.
\end{equation*}
If $R(\gamma)$ is to preserve at least $\frac{31}{32}$ of the
supersymmetry, then $\codim(K^0\subset K) \leq 1$, whence
$\codim(K_-^0 \subset K_-) \leq 1$.  Now let $\psi_- \in K^0_-$.  We
have that $g_S \psi_- = \psi_-$ and $g_N \psi_- = \psi_-$.  In
particular, the space of $g_S$-invariants in $K_-$ must have
codimension at most $1$: it is either $15$- or $16$-dimensional.  We
claim that this means that $g_S =1$.  Indeed, $g_S$ is obtained by
exponentiating the semisimple part of $R(X)$:
\begin{align*}
  g_S &= \exp \left(v^- R(\be_-) + \theta^1 R(\bM_{12}) + \theta^2
    R(\bM_{45}) + \theta^3 R(\bM_{67}) + \theta^4
    R(\bM_{89})\right)\\
  &= e^{-\frac{\mu v^-}{4}I\Pi_+} e^{-\frac{\mu v^-}{12}I\Pi_-}
  e^{\frac{\theta^1}2 \Gamma_{12}}   e^{\frac{\theta^2}2 \Gamma_{45}}
  e^{\frac{\theta^3}2 \Gamma_{67}}   e^{\frac{\theta^4}2 \Gamma_{89}}~,
\end{align*}
whose action on $\psi_- \in K_-$ is given by
\begin{align*}
  g_S \psi_- = e^{-\frac{\mu v^-}{12}I} e^{\frac{\theta^1}2
    \Gamma_{12}}   e^{\frac{\theta^2}2 \Gamma_{45}}
  e^{\frac{\theta^3}2 \Gamma_{67}}   e^{\frac{\theta^4}2
    \Gamma_{89}} \psi_-~.
\end{align*}
But now notice that each of the factors in $g_S$ is of the form
$e^{\half\theta^k J_k}$ for commuting complex structures $J_k$, which
can be simultaneously diagonalised upon complexifying $K_-$.  This is
precisely the set-up in §\ref{sec:even-multiplicity}, with $W = K_-
\otimes_\RR \CC$ and $N=5$.  Therefore we may apply the
even-multiplicity argument to conclude that the space of such $\psi_-$
is always divisible by $2$, whence $g_S$ cannot preserve exactly $15$
such spinors and must in fact preserve all $16$.

This means that $g_S = 1$, whence $\theta^i \in 4\pi\ZZ$ and $\mu v^-
\in 24 \pi \ZZ$.  The condition on the $\theta^i$ say that this part
of the group element is trivial, whence the semisimple part of the
group element $\gamma$ is given by
\begin{equation*}
  \gamma_S = \exp\left(\frac{24\pi k}{\mu} \be_-\right) \qquad
  \text{for some $k\in\ZZ$},
\end{equation*}
which acts trivially on the Killing spinors.  In addition, it follows
from equation~\eqref{eq:expad} that this element belongs to the kernel
of the adjoint representation and hence to the centre of $S$.

It remains to show that the nilpotent part of $\gamma$ cannot preserve
precisely a fraction $\frac{31}{32}$ of the supersymmetry.   Since
$\gamma_S$ is central, we have that $g_N = \1 -\half \alpha \Gamma_+$,
for an endomorphism $\alpha$ given by
\begin{equation}
  \label{eq:alpha}
  \alpha = \sum_{i=1}^9 \left(\lambda_i^2 v^i \Gamma_i + \lambda_i w^i I
    \Gamma_i \right)~,
\end{equation}
where the coefficients $v^i$, $w^i$ are the ones appearing in the
expression for $X\in\fg$ in equations (\ref{eq:X1}) and
(\ref{eq:X2}).  It is clear from the form of $g_N$ that it acts like
the identity on $K_+$ and that the equation $g_N \psi = \psi$ becomes
\begin{equation*}
  (\1 -\half \alpha \Gamma_+)(\psi_+ + \psi_-) = \psi_+ + \psi_- \implies
  \alpha \Gamma_+\psi_- = 0~.
\end{equation*}
Since $\Gamma_+$ has no kernel on $K_-$, it follows that we must
investigate the kernel of $\alpha$ on $K_+$ or, defining
$\check\alpha$ by $\alpha\Gamma_+ = \Gamma_+\check\alpha$, the kernel
of
\begin{equation*}
  \check\alpha = \sum_{i=1}^9 \left(\lambda_i^2 v^i \Gamma_i -
    \lambda_i w^i I \Gamma_i \right)
\end{equation*}
on $K_-$.
% To show that the kernel cannot have codimension $1$, it is
% equivalent to show that $\check\alpha$ cannot have rank exactly $1$.
As discussed above, there are two cases we must consider,
corresponding to the forms (\ref{eq:X1}) and (\ref{eq:X2}) for $X$.

For $X$ given by equation (\ref{eq:X1}), the coefficients $w^i=0$ and
hence $\check\alpha = \sum_{i=1}^9 \lambda_i^2 v^i \Gamma_i$ is given
by the Clifford product by a vector with components $(\lambda_i^2
v_i)$ and by the Clifford relations, provided the vector is nonzero,
this endomorphism has trivial kernel.  If the vector is zero, so that
$v^i=0$, then we preserve all the supersymmetry.  In summary, in this
case we may quotient by a subgroup of the centre generated by
$\exp(v^+\be_+)$, for some $v^+$, while preserving all of the
supersymmetry.

For $X$ given by equation (\ref{eq:X2}), we can argue as follows.
First of all, since we are already in the situation when $g_S = 1$, we
have some more freedom in choosing the normal form.  In particular, we
may conjugate by $\SO(3) \times \SO(6)$ to set all $w^i=0$ except for
$w^1$ and $w^4$.  This still leaves an $\SO(2) \times \SO(5)$ which
can be used to set $v^{3,6,7,8,9}=0$.  Finally we may conjugate by
$e^{t\be_-}$ to set $w^4=0$, which then gives the further freedom
under $\SO(6)$ to set $v^5=0$.  In summary, and after relabeling,
we remain with
\begin{equation*}
  X = v^+\be_+ + v^1 \be_1^* + v^2 \be^*_2 + v^4 \be^*_4 + w^1\be_1~.
\end{equation*}
A calculation (performed on computer) shows that the characteristic
polynomial of the endomorphism $\check\alpha$ has the form
\begin{equation*}
  \chi_{\check\alpha}(t) = (t^4 + 2 A t^2 + B)^4 =
  \mu_{\check\alpha}(t)^4~,
\end{equation*}
where the notation is such that $\mu_{\check\alpha}$ is the minimal
polynomial, and $A$ and $B$ are given in terms of $\bz =
(v^1,v^2,v^4,w^1)^t$ and $\bz^2 = ((v_1)^2,(v^2)^2,(v^4)^2,(w^1)^2)^t$
as
\begin{equation*}
  A = |\bz|^2 \qquad\text{and}\qquad B = Q(\bz^2)~,
\end{equation*}
where $Q$ is the quadratic form defined by the matrix
\begin{equation*}
  \begin{pmatrix}
    1 & -1 & 1 & -1 \\
    -1 & 1 & 1 & 1 \\
    1 & 1 & 1 & 1\\
    -1 & 1 & 1 & 1
  \end{pmatrix}~.
\end{equation*}
This matrix is not positive-definite: it has eigenvalues
$0,2,1\pm\sqrt5$ with respective eigenvectors
\begin{equation*}
  \begin{pmatrix}
    0\\-1\\0\\1
  \end{pmatrix}\quad
  \begin{pmatrix}
    1\\0\\1\\0
  \end{pmatrix}\quad
  \begin{pmatrix}
    \half(1\mp \sqrt5)\\1\\-\half(1\mp \sqrt5)\\1
  \end{pmatrix}~.
\end{equation*}
Clearly, $\check\alpha$ will have nontrivial kernel if and only if
$B=0$, which means $Q(\bz^2) = 0$. Taking into
account that the components of the vector $\bz^2$ are non-negative, we
see that this implies that $v^2=0$ and $(v^1)^2 + (v^4)^2 = (w^1)^2$.
The characteristic polynomial again satisfies $\chi_{\check\alpha}(t)
= \mu_{\check\alpha}(t)^4$, where now
\begin{equation*}
  \mu_{\check\alpha}(t) = t^2(t^2 + 4 (w^1)^2)~.
\end{equation*}
The dimension of the kernel of $\check\alpha$ will be less than the
algebraic multiplicity of $0$ as a root of the characteristic
polynomial. Therefore for $w^1\neq 0$, the dimension of the kernel
will be at most $8$ (and, in fact, it is exactly $8$).  For $w^1=0$,
the endomorphism $\check\alpha \equiv 0$ and the dimension of the
kernel is precisely $16$.

In summary, we have shown that if an element in the image of the
exponential map of the symmetry groups of the maximally supersymmetric 
vacua preserves at least $31$ supersymmetries, it must in fact
preserve all $32$.

\section{Investigating the exponential map}
\label{sec:exp}

As outlined in the introduction, to complete the proof of the
non-existence of preonic quotients, it is crucial to show that every
element in the symmetry group of a maximally supersymmetric background
has finite index.  In other words, that for every element $\gamma \in
G$, there is a positive integer $n$ (which may depend on $\gamma$)
such that $\gamma^n$ lies in the image of the exponential map.  This
is a weaker condition than surjectivity of the exponential map.  We
remind the reader that a Lie group is said to be \textbf{exponential}
if the exponential map is surjective.  It is plain to see that if a
simply-connected Lie group is exponential, then so are all connected
Lie groups sharing the same Lie algebra.  We will find the following
partial converse result very useful.

\begin{lemma}
  Let $\pi:\widehat G \to G$ be a finite cover; that is, a surjective
  homomorphism with finite kernel, with $\widehat G$ (and hence $G$)
  connected.  Then if every element of $G$ has finite index (of
  exponentiality), so does every element of $\widehat G$.
\end{lemma}

\begin{proof}
  Let $\fg$ be the Lie algebra of both $G$ and $\widehat G$ and let
  $\exp: \fg \to G$ and $\widehat{\exp}: \fg \to \widehat G$ denote
  the corresponding exponential maps, related by the following
  commutative diagramme:
  \begin{equation*}
    \xymatrix{& \widehat G \ar[d]^\pi \\
      \fg \ar[ur]^{\widehat\exp} \ar[r]^{\exp} & G}
  \end{equation*}
  Since $\pi: \widehat G \to G$ is a finite cover, $Z = \ker \pi$ is a
  finite subgroup of the centre of $\widehat G$.  (This is because
  a normal discrete subgroup of a connected Lie group is central.)  Now let
  $\widehat\gamma \in \widehat G$.  Since $\gamma = \pi(\widehat
  \gamma)$ has finite index, there exists some
  positive integer $N$ such that $\gamma^N= \exp(X)$ for some $X \in
  \fg$.  Since
  \begin{equation*}
    \pi(\widehat\gamma^N) = \pi(\widehat\gamma)^N = \gamma^N = \exp(X)
    = \pi(\widehat{\exp}(X))~,
  \end{equation*}
  it follows that there is some $z \in Z$ for which $\widehat\gamma^N
  = z\widehat{\exp}(X)$.  Since $Z$ is finite, $z$ has finite order,
  say, $|z|$, whence
  \begin{equation*}
    \widehat\gamma^{N|z|} = \widehat{\exp}(X)^{|z|} =
    \widehat{\exp}(|z|X)~,
  \end{equation*}
  and $\widehat\gamma$ also has finite index.
\end{proof}

By virtue of this lemma it is sufficient to exhibit a finite quotient
(by a central subgroup) of the groups under consideration for which
every element has finite index.

Let us consider the exponential properties of the symmetry groups $G$
of the maximally supersymmetric backgrounds, or more precisely of the
related groups which act effectively on the Killing spinors.

The flat backgrounds have symmetry groups $\SO(1,d-1) \ltimes
\RR^{1,d}$, but as translations act trivially on spinors, it is only
$\Spin(1,d-1)$ which concerns us.  It follows, for example, from the
classification results of \cite{DjoNgu} that $\Spin(1,2n)$ for $n\geq
2$ and $\Spin(1,2m-1)$ for $m\geq 3$ are indeed exponential.

The Freund--Rubin backgrounds have symmetry groups
$\widetilde{\SO(2,p)} \times \SO(q)$ for various values of $p$ and
$q$, where $\widetilde{\SO(2,p)}$ is the universal covering group.  As
discussed, for example, in \cite{FigSimAdS} the groups acting
effectively on the Killing spinors are $\Spin(2,p) \times \Spin(q)$.
Being compact, $\Spin(q)$ is exponential, whereas $\Spin(2,p)$ is not
exponential, but nevertheless, as reviewed in \cite{Exponential}, it
follows from the results in \cite{Sibuya} that the square of every
element is in the image of the exponential map.

Finally, the symmetry group of the maximally supersymmetric wave has
the form $G \rtimes S$ where $G$ is the solvable transvection group of
an $11$-dimensional lorentzian symmetric space \cite{CahenWallach} and
$S = \SO(3) \times \SO(6) < \SO(9)$ is a subgroup of the transverse
rotation group leaving invariant the fluxes and the matrix $A$
defining the metric.  Since $S$ is compact its spin cover, which is
the group acting on the Killing spinors, is actually exponential,
whence we need only concentrate on the group $G$.

To understand this case better it pays to look at a toy model.  Let
$\fg$ denote the four-dimensional Lie algebra with basis
$(\be_\pm,\be,\be^*)$ and nonzero brackets
\begin{equation*}
  [\be_-,\be] = \be^* \qquad [\be_-,\be^*] = -\be \qquad [\be,\be^*] =
  \be_+~.
\end{equation*}
This is the extension of the Heisenberg subalgebra spanned by
$(\be,\be^*,\be_+)$ by the outer derivation $\be_-$ which acts by
infinitesimal rotations in the $(\be,\be^*)$ plane.  It is also known
as the Nappi-Witten algebra \cite{NW}.  It is a Lie subalgebra of
$\fgl(4,\RR)$.  Indeed, a possible embedding is given by
\begin{equation}
  \label{eq:toyalg}
  x \be^* + y \be + z\be_+ + t \be_- \mapsto
  \begin{pmatrix}
    0 & t & x & 0\\
    -t & 0 & y & 0\\
    0 & 0 & 0 & 0\\
    -y & x & -2z & 0
  \end{pmatrix}~,
\end{equation}
and the Lie subgroup $G < \GL(4,\RR)$ with this Lie algebra is given
by
\begin{equation*}
  G = \left\{
    \begin{pmatrix}
      \cos t & \sin t & x & 0 \\
      -\sin t & \cos t & y & 0 \\
      0 & 0 & 1 & 0\\
      -y\cos t - x \sin t & x \cos t - y \sin t & - 2 z & 1
    \end{pmatrix} \middle | x,y,z,t\in\RR \right\}~,
\end{equation*}
which is diffeomorphic to $S^1 \times \RR^3$.  Its universal cover
$\widetilde G$ is diffeomorphic to $\RR^4$ and is obtained by
unwinding the circle; that is, removing the periodicity of the
$t$-coordinate.  Let $g(x,y,z,t)$ denote the group element with
coordinates $(x,y,z,t) \in \RR^4$.  It is convenient, as in
\cite[§3.1]{FSNW} but using a different notation, to introduce the
complex variable $w= x+iy$ in terms of which the group multiplication
on $\widetilde G$ is given explicitly by
\begin{equation*}
  g(w_1,z_1,t_1) g(w_2,z_2,t_2) = g(w',z',t')~,
\end{equation*}
where
\begin{align*}
  t' &= t_1 + t_2 \\
  w' &= w_1 + e^{-it_1} w_2\\
  z' &= z_1 + z_2 - \half \im \left(\overline w_1 e^{-it_1} w_2\right)~.
\end{align*}

The element $g(0,0,t) = \exp(t\be_-)$ is central and generates the
action of the fundamental group of $G$ on the universal cover
$\widetilde G$.  Any representation of $\widetilde G$ for which
$\exp(t\be_-)$ acts with period $2\pi n$, for some integer $n$, will
factor through an $n$-fold cover $\widehat G$ of $G$, namely the
quotient of $\widetilde G$ by the infinite cyclic subgroup of the
centre generated by $g(0,0,2\pi n)$.  It follows from the Lemma and
the fact, to be proven shortly, that $G$ is exponential, that every
element of $\widehat G$ has finite index.

We now show that that $G$ is exponential.  Indeed, the matrix in
\eqref{eq:toyalg} exponentiates inside $\GL(4,\RR)$ to
\begin{equation*}
  \begin{pmatrix}
    \cos t & \sin t & \frac{x \sin t - y (\cos t -1)}{t} & 0\\
    -\sin t & \cos t & \frac{x (\cos t - 1) + y \sin t}{t} & 0\\
    0 & 0 & 1 & 0\\
    -\frac{-y \sin t + x (\cos t - 1)}{t} & \frac{x \sin t + y (\cos t
      -1)}{t} & -2z + \frac{\sin t -t}{t^2}(x^2+y^2) & 1
  \end{pmatrix}~.
\end{equation*}
It is clear that this is surjective provided that the linear
transformation
\begin{equation*}
  \begin{pmatrix}
    x \\ y 
  \end{pmatrix}
  \mapsto
  \begin{pmatrix}
    \frac{x \sin t - y (\cos t -1)}{t}\\
    \frac{x (\cos t - 1) + y \sin t}{t}
  \end{pmatrix}
 =
 \frac1t
 \begin{pmatrix}
   \sin t & 1 - \cos t\\
   \cos t - 1 & \sin t
 \end{pmatrix}
  \begin{pmatrix}
    x \\ y 
  \end{pmatrix}
\end{equation*}
is nonsingular.  Now the above linear transformation is singular
precisely when $t \in 2\pi\ZZ$, but $t \neq 0$.  However the 
group elements
\begin{equation*}
  \begin{pmatrix}
    1 & 0 & x & 0 \\
    0 & 1 & y & 0 \\
    0 & 0 & 1 & 0 \\
    -y & x & -2z & 1
  \end{pmatrix}
\end{equation*}
with such a value of $t$, coincide with those with $t=0$, for which
the above transformation is nonsingular and are hence in the image of
the exponential map; explicitly,
\begin{equation*}
  \begin{pmatrix}
    1 & 0 & x & 0 \\
    0 & 1 & y & 0 \\
    0 & 0 & 1 & 0 \\
    -y & x & -2z & 1
  \end{pmatrix} = \exp
  \begin{pmatrix}
    0 & 0 & x & 0\\
    0 & 0 & y & 0\\
    0 & 0 & 0 & 0\\
    -y & x & -2z & 0
  \end{pmatrix}~.
\end{equation*}
Therefore we conclude that the group $G$ is exponential.

The transvection group of a Cahen--Wallach space is a more complicated
version of the Nappi--Witten group.  The Lie algebra is spanned by
$(\be_\pm,\be_i,\be_i^*)$ for $i=1,\dots,d-2$ with nonzero brackets
\begin{equation*}
  [\be_-,\be_i]=\be_i^* \qquad [\be_-,\be_i^*] = A_{ij}\be_j\qquad
  [\be_i^*,\be_j]=A_{ij}\be_+~,
\end{equation*}
for some non-degenerate symmetric matrix $A_{ij}$.  Although a more
general analysis is indeed possible, we shall concentrate uniquely on
those matrices $A$ which are negative-definite.  Without loss of
generality we can rewrite the Lie algebra as
\begin{equation*}
  [\be_-,\be_i]=\be_i^* \qquad [\be_-,\be_i^*] = -\lambda_i^2 \be_i\qquad
  [\be_i^*,\be_j]=-\lambda_i^2 \delta_{ij} \be_+~,
\end{equation*}
for some $\lambda_i>0$.  It is convenient to change basis $\be_i \mapsto
\lambda_i^{-1/2} \be_i$ and $\be^*_i \mapsto \lambda_i^{-3/2}
\be^*_i$, relative to which the brackets now take a more symmetrical
form
\begin{equation*}
  [\be_-,\be_i]=\lambda_i \be_i^* \qquad [\be_-,\be_i^*] = -\lambda_i
  \be_i\qquad [\be_i^*,\be_j]=- \delta_{ij} \be_+~.
\end{equation*}
We exhibit this Lie algebra as a subalgebra of $\fgl(2d-2,\RR)$ via
the following embedding
\begin{equation}
  \label{eq:CWliealg}  
 \sum_{i=1}^{d-2}( x_i\be_i^* + y_i\be_i) +  t\be_- + z \be_+ \mapsto
 \begin{pmatrix}
   0 & \lambda_1 t & & & & x_1 & 0\\
   -\lambda_1 t & 0 & & & & y_1 & 0\\
   & & \ddots & & & \vdots & \vdots \\
   & & & 0 & \lambda_{d-2} t & x_{d-2} & 0\\
   & & & -\lambda_{d-2} t & 0 & y_{d-2} & 0\\
   0 & 0 & \cdots & 0 & 0 & 0 & 0 \\
   -y_1 & x_1 & \cdots & -y_{d-2} & x_{d-2} & -2 z & 0
 \end{pmatrix}~.
\end{equation}
The Lie subgroup $G < \GL(2d-2,\RR)$ with this Lie algebra consists of
matrices of the form
\begin{equation*}
  \begin{pmatrix}
   \cos \lambda_1 t & \sin \lambda_1 t & & & & x_1 & 0\\
   -\sin \lambda_1 t & \cos \lambda_1 t & & & & y_1 & 0\\
   & & \ddots & & & \vdots & \vdots \\
   & & & \cos\lambda_{d-2}t & \sin \lambda_{d-2} t & x_{d-2} & 0\\
   & & & -\sin \lambda_{d-2} t & \cos\lambda_{d-2}t & y_{d-2} & 0\\
   0 & 0 & \cdots & 0 & 0 & 1 & 0 \\
   u_1 & v_1 & \cdots & u_{d-2} & v_{d-2} & -2 z & 1
  \end{pmatrix}
\end{equation*}
where
\begin{equation*}
  u_i = -y_i \cos\lambda_i t - x_i \sin\lambda_i t
  \qquad
  v_i = x_i \cos\lambda_i t - y_i \sin\lambda_i t~.
\end{equation*}
The exponential of the matrix in equation \eqref{eq:CWliealg} is given
by
\begin{equation*}
  \begin{pmatrix}
   \cos \lambda_1 t & \sin \lambda_1 t & & & & X_1 & 0\\
   -\sin \lambda_1 t & \cos \lambda_1 t & & & & Y_1 & 0\\
   & & \ddots & & & \vdots & \vdots \\
   & & & \cos\lambda_{d-2}t & \sin \lambda_{d-2} t & X_{d-2} & 0\\
   & & & -\sin \lambda_{d-2} t & \cos\lambda_{d-2}t & Y_{d-2} & 0\\
   0 & 0 & \cdots & 0 & 0 & 1 & 0 \\
   U_1 & V_1 & \cdots & U_{d-2} & V_{d-2} & -2 Z & 1
  \end{pmatrix}
\end{equation*}
where
\begin{equation}
  \label{eq:xy2XY}
  \begin{aligned}[t]
    X_i &= \frac{x_i \sin\lambda_i t + y_i (1 - \cos\lambda_it)}{\lambda_i t}\\
    Y_i &= \frac{y_i \sin\lambda_i t - x_i (1 - \cos\lambda_it)}{\lambda_i t}\\
    U_i &= -Y_i \cos\lambda_i t - X_i \sin\lambda_i t\\
    V_i &= X_i \cos\lambda_i t - Y_i \sin\lambda_i t\\
    Z &= z - \half \sum_{i=1}^{d-2} (x_i^2 + y_i^2) \frac{(\lambda_i t
      - \sin\lambda_i t)}{\lambda_i^2 t^2}~.
  \end{aligned}
\end{equation}

We will now specialise further to the geometry of interest, where the
$\lambda_i$ are given by equation \eqref{eq:CW11}.  It is important to
observe that the ratios of the $\lambda_i$ are rational --- in fact,
integral.  This means that whereas the group is not exponential, as we
will now see, nevertheless the square of every element lies in the
image of the exponential map.

We will now take the $\lambda_i$ given by equation \eqref{eq:CW11}.
The surjectivity of the exponential map is only in question when the
linear map from $(x_i,y_i)$ to $(X_i,Y_i)$ in equation
\eqref{eq:xy2XY} fails to be an isomorphism.  This happens whenever
$\lambda_i t \in 2\pi\ZZ$ and $t\neq 0$.  For the $\lambda_i$ under
consideration, this happens whenever $\mu t \in 6 \pi \ZZ$, but $\mu
t \neq 0$.  Let $\mu t = 6 \pi n$ and $n\neq 0$.  Then the group
elements with such values of $t$ are given by
\begin{equation*}
  \begin{pmatrix}
   1 & 0 & & & & x_1 & 0\\
   0 & 1 & & & & y_1 & 0\\
   & & \ddots & & & \vdots & \vdots \\
   & & & (-1)^n & 0 & x_9 & 0\\
   & & & 0 & (-1)^n & y_9 & 0\\
   0 & 0 & \cdots & 0 & 0 & 1 & 0 \\
   -y_1 & x_1 & \cdots & -(-1)^n y_9 & (-1)^n x_9 & -2 z & 1
  \end{pmatrix}~,
\end{equation*}
whence we see that if $n$ is even, this is the same as if $t=0$ for
which the exponential map is surjective, whereas if $n$ is odd, then
this is not in the image of the exponential map, but its square
is again of the form of the matrices with $t=0$ and hence in the image
of the exponential map.  In other words, for every $g\in G$, $g^2 \in
E_G$.

Finally we observed above that the action of $\be_-$ on the Killing
spinors is such that $\exp(t\be_-)$ is periodic, whence the group
$\widehat G$ acting effectively on spinors is a finite cover of the
matrix group $G$.  By the Lemma and the results above, every element
in $\widehat G$ has finite index.

\section{Other supergravity theories}
\label{sec:other}

The results in this paper complete the proof of \cite{NoMPreons}
of the non-existence of preonic M-theory backgrounds.  How about other
supergravity theories?

In \cite{NoIIBPreons}, it was shown that any solution of IIB
supergravity preserving 31 supersymmetries is locally maximally
supersymmetric.  In the IIB case, this result excludes the possibility
of obtaining preons by quotients, because the symmetry groups of the
maximally supersymmetric vacua of IIB act complex linearly on Killing
spinors, in the conventions where the spinors in IIB are complex
chiral spinors.  Hence the space of invariants must be a complex
subspace and hence must have even dimension.

For IIA supergravity, the result follows from the one for
eleven-dimensional supergravity.  Indeed, any supergravity background
preserving $31$ supersymmetries will oxidise to an eleven-dimensional
supergravity background preserving at least as much supersymmetry,
which by the results of \cite{NoMPreons} and of the present paper,
must in fact be maximally supersymmetric.  Furthermore, the IIA
background is then a quotient of this eleven-dimensional background by
a monoparametric subgroup of symmetries, but as we have shown in this
paper, no element in that group preserves exactly $31$
supersymmetries, whence if it preserves at least $31$ it must preserve
them all.

The same argument applies to any other supergravity theory with $32$
supercharges obtained by dimensional reduction of $d{=}11$ or IIB
supergravities.  In summary, there are no supergravity backgrounds, in
any known supergravity theory, preserving a fraction $\frac{31}{32}$
of the supersymmetry.

\section*{Acknowledgments}

It is a pleasure to thank George Papadopoulos for prompting us to look
into this problem, Dmitri Alekseevsky for very useful discussions and
Karl Hofmann for guiding us through the more recent literature on the
exponential map.

\bibliographystyle{utphys}
\bibliography{AdS,AdS3,ESYM,Sugra,Geometry}

\end{document}